\documentclass[a4paper,11pt]{article}

\usepackage{contribution}

\newcommand{\weblink}[2][]{%
    \ifthenelse{\equal{#1}{}}%
    {\textnormal{\url{#2}}}%
    {\textnormal{\href{#2}{#1}}}%
}

\def\beq{\begin{equation}}
\def\eeq#1{\label{#1}\end{equation}}
\def\eeqn{\end{equation}}

\def\beqa{\begin{eqnarray}}
\def\eeqa#1{\label{#1}\end{eqnarray}}
\def\eeqan{\end{eqnarray}}

\let\bar=\overbar

\def\Dslash{\not{\hbox{\kern-4pt $D$}}}
\def\dslash{\not{\hbox{\kern-2pt $\del$}}}

\def\msb{{\bar{\ssstyle M \kern -1pt S}}}

\newcommand{\contribution}[7][]{%
  \clearpage
  \thispagestyle{plain}
  \ifthenelse{\equal{#1}{}}
  {\hypersetup{pdftitle={#2}}}
  {\hypersetup{pdftitle={#1}}}
  \hypersetup{pdfauthor={{#3} {#4}}}
  {\centering\normalfont\LARGE\bfseries\sffamily #2 \par\nobreak}
  \lhead{}
  \chead{%
    \textit{\footnotesize XIV International Conference on Hadron Spectroscopy
      (\weblink[\textit{hadron2011}]{http://www.hadron2011.de}), 13-17 June 2011, Munich, Germany}%
  }
  \rhead{}
  \bigskip
  \begin{center}
    {#3} {#4}\ifthenelse{\equal{#6}{}}{}{\footnote{\weblink[#6]{mailto:#6}}}
    \ifthenelse{\equal{#7}{}}{}{#7} \\
    \textit{#5}
  \end{center}
  \bigskip
}

\renewcommand{\abstract}[1]{%
  \begin{center}
    \begin{minipage}{0.85\textwidth}
      \begin{footnotesize}
        #1
      \end{footnotesize}
    \end{minipage}
  \end{center}
  \bigskip
}

\begin{document}

%
%
%
%
%
{  


\newcommand{\Erw}[1]{\left\langle #1 \right\rangle}
\newcommand{\Ord}{\mathcal{O}}
%

\contribution[Electromagnetic form factor of the pion from lattice QCD]
{Calculation of the pion electromagnetic form factor from lattice QCD} 
{Bastian B.}{Brandt$^\ast$}  
{$^\ast$ Institut f\"ur Kernphysik,
  University of Mainz,
  Becher-Weg 45, D-55099\\
 $^\dagger$ CERN, Physiks Department, TH Unit, CH-1211 Geneva 23\\
 $^\ddagger$ Helmholtz Institute Mainz, University of Mainz, D-55099 Mainz}  %
{brandt@kph.uni-mainz.de}  
{, Andreas J\"uttner$^\dagger$, Hartmut Wittig$^{\ast\ddagger}$}  %
%

\abstract{%
  We present a lattice calculation of the vector form factor of the pion for two
flavours of non-perturbatively $O(a)$ improved Wilson fermions. For the
measurements we utilise the CLS ensembles which include various lattice spacings
and pion masses down to about 250 MeV. To obtain a fine momentum resolution
near zero momentum transfer ($q^2$) partially twisted boundary conditions are
employed using several twist angles. Due to the fine resolution around $q^2=0$
we are able to determine the slope of the form factor and, in turn, extract the
charge radius of the pion without any model dependence. The results for the form
factor and the charge radius are then compared to chiral perturbation theory and
phenomenological models which are used to extrapolate the results to the
physical point.
}
%


Over the last years Monte-Carlo simulations of lattice QCD
started to produce accurate and reliable results for a number of quantities of
phenomenological interest, such as e.g. the spectrum of the low lying hadrons
and light quark masses (see e.g. \cite{Durr:2008zz} and
\cite{Colangelo:2010et}). Despite good agreement between theory and experiment
for these quantities there are others where lattice QCD does not coincide with
experiment. The origin of these discrepancies is not clear since a number of
systematic effects have to be controlled both in experiment and simulations. A
particular example where experiment and theory are not in satisfactory
agreement are observables connected with structural properties of the nucleon,
such as electric and magnetic form factors as well as the nucleon axial charge
(see e.g.
\cite{Alexandrou:2010cm,Collins:2011mk,Capitani:2010sg,Brandt:2011sj}).
Another technically simpler observable where similar systematic effects enter is
the pion electromagnetic form factor $f_{\pi\pi}(q^2)$, where $q$ is the
momentum transfer. The fact that $f_{\pi\pi}$ receives no contribution from
quark-disconnected diagrams for two degenerate flavours makes it the
ideal observable to perform a precision calculation in lattice QCD.
Nevertheless, in the region of small momentum transfers the extraction of
$f_{\pi\pi}(q^2)$ usually suffers from an intrinsic model dependence, since a
direct calculation in that region has not been possible so far for both
experiment and theory. Related to the pion form factor in that kinematical
regime is the pion charge radius $\Erw{r_\pi^2}$ defined by the linear behavior
of $f_{\pi\pi}(q^2)$ at $q^2=0$,
\begin{equation}
f_{\pi\pi}(q^2) = 1 + \frac{\Erw{r_\pi^2}}{6} \: q^2 + \ldots
 \quad \Rightarrow \quad
\left.\Erw{r_\pi^2} =
6\:\frac{d\:f_\pi(q^2)}{dq^2}\right|_{q^2=0} \;.
\end{equation}
The extraction of $\Erw{r_\pi^2}$ is thus mostly governed by the modeling of
the $q^2$-dependence of $f_{\pi\pi}$ unless results are available around
$q^2\approx0$. In lattice QCD the accessible momenta are usually obtained by
Fourier transformation and thus constrained by finite lattice volume. This has
changed recently by the introduction of partially twisted boundary conditions
\cite{Bedaque:2004kc,deDivitiis:2004kq} that in principle allow arbitrary small
momentum transfers in lattice simulations \cite{Boyle:2007wg}.

\begin{table}[t]
\centering
\vspace*{-3mm}
\begin{tabular}{ccccccc}
\hline
\hline
$\beta$ & $a$[fm] & lattice & \# masses & $m_\pi\:L$ & Labels & Statistic \\
\hline
5.20 & 0.08 & $64\times32^3$ & 3 & 6.0 -- 4.0 & A3 -- A5 & $\Ord(100)$ \\
\hline
5.30 & 0.07 & $64\times32^3$ & 2 & 6.2, 4.7 & E4, E5 & $\Ord(100)$ \\
5.30 & 0.07 & $96\times48^3$ & 1 & 5.0 & F6 & 233 \\
\hline
5.50 & 0.05 & $96\times48^3$ & 3 & 7.7 -- 5.3 & N3 -- N5 & $\Ord(100)$ \\
\hline
\hline
\end{tabular}
\caption{Compilation of simulation parameters.}
\vspace*{-3mm}
\label{tab1}
\end{table}

Our calculation of $f_{\pi\pi}$ employs ensembles generated in the context of
the CLS project\footnote{{\tt https://twiki.cern.ch/twiki/bin/view/CLS/WebHome}}
and include three different lattice spacings with three different pion masses
each. The main parameters of the ensembles used in the analysis are shown in
table \ref{tab1}. To reduce the statistical noise we use stochastic $Z(2)\times
Z(2)$ wall sources for the computation of the quark propagators, see e.g.
\cite{Boyle:2008rh}. The momenta are generated by five twist angles tuned so as
to obtain as many as 30 values of $q^2$ below the lowest accessible $q^2$ from
Fourier-momentum. We express our result in units of the Sommer scale $r_0$
\cite{Sommer:1993ce} in the chiral limit, which has been measured for these
ensembles in \cite{Donnellan:2010mx}. To compare our data to experimental
results and results from other collaborations we use $r_0=0.471$~fm as obtained
in \cite{hip_latt2011}. To make optimal use of the generated data we extract
$f_{\pi\pi}$ using a combination of the three different ratios of two- and
three-point functions defined in \cite{Boyle:2007wg}. The error bars are
estimated with the bootstrap method using 1000 samples. For more details of the
simulations see \cite{Brandt:2010ed} and \cite{Brandt:2011sj} as well as our
upcoming publication.


\begin{figure}[t]
 \centering
\vspace*{-3mm}
\includegraphics[angle=-90, width=0.45\textwidth]{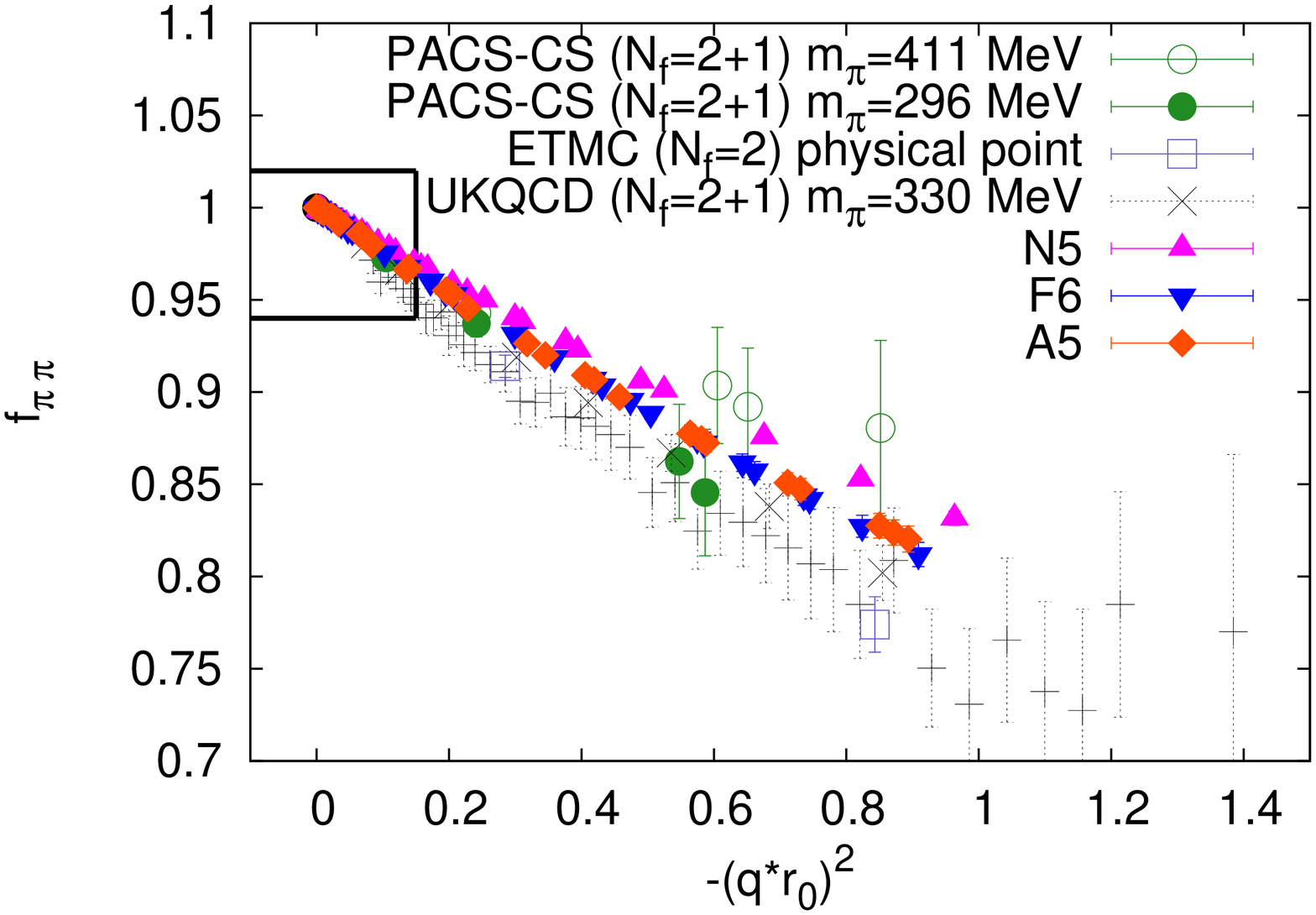} \hfil
\includegraphics[angle=-90, width=0.45\textwidth]{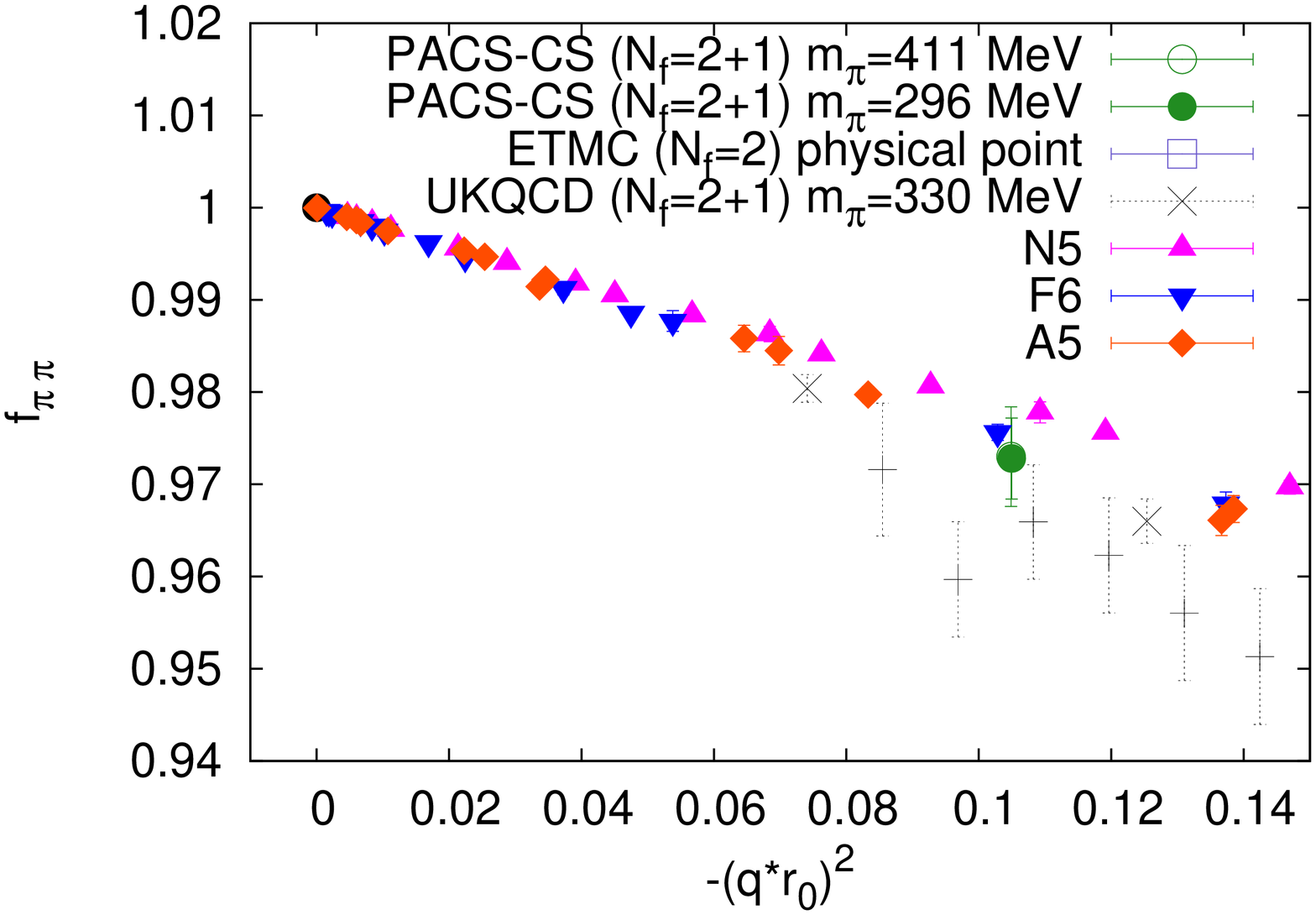}
\vspace*{-3mm}
\caption{Results for the pion form factor for the lightest quark mass for each
lattice spacing compared with the results from PACS-CS \cite{Nguyen:2011ek},
ETMC \cite{Frezzotti:2008dr} and UKQCD \cite{Boyle:2008yd}, as well as the
experimental results from \cite{Amendolia:1986wj}. The right figure is the inset
in the top left-hand corner.}
\label{fig1}
\vspace*{-3mm}
%
\includegraphics[angle=-90, width=0.45\textwidth]{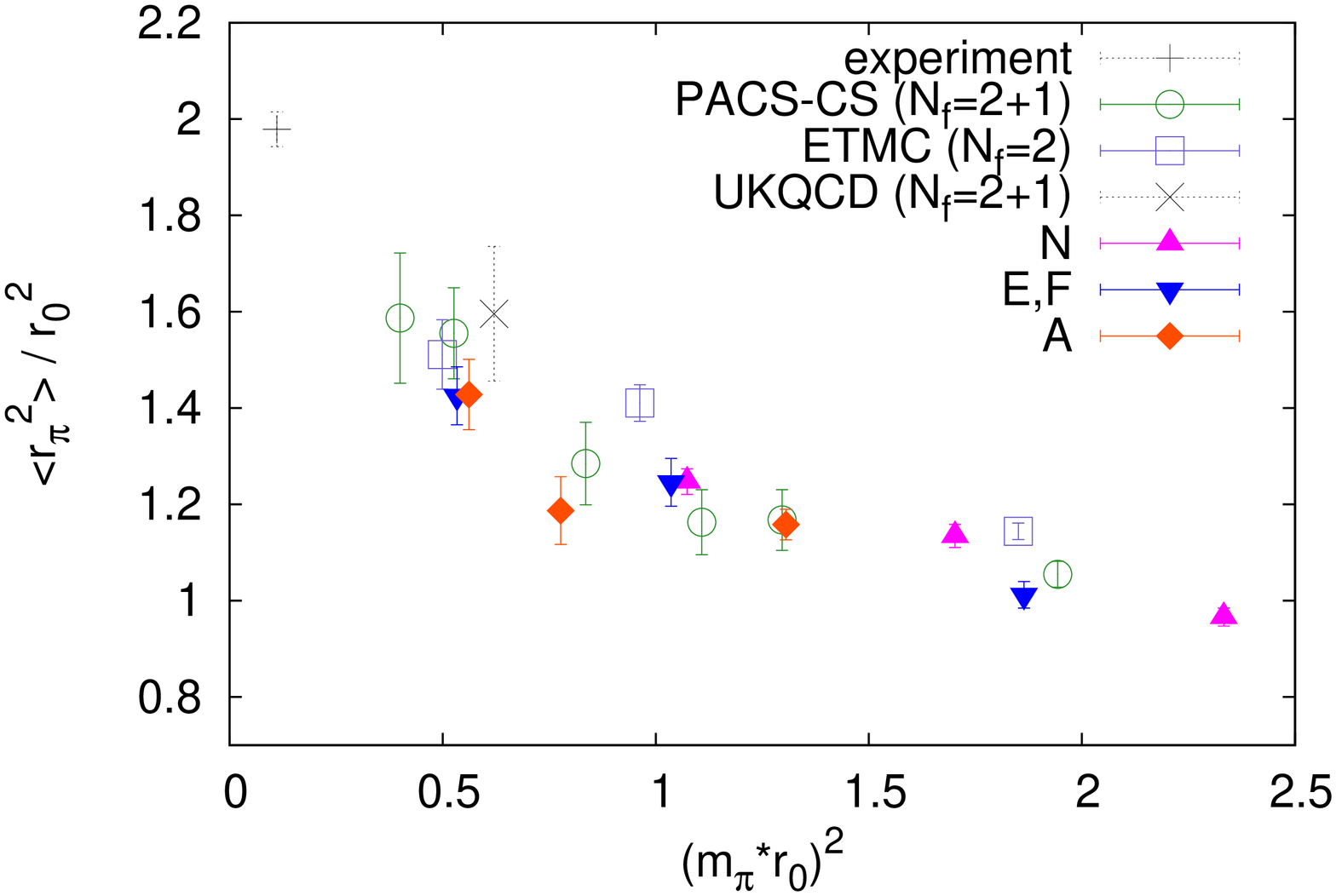} \hfil
\includegraphics[angle=-90, width=0.45\textwidth]{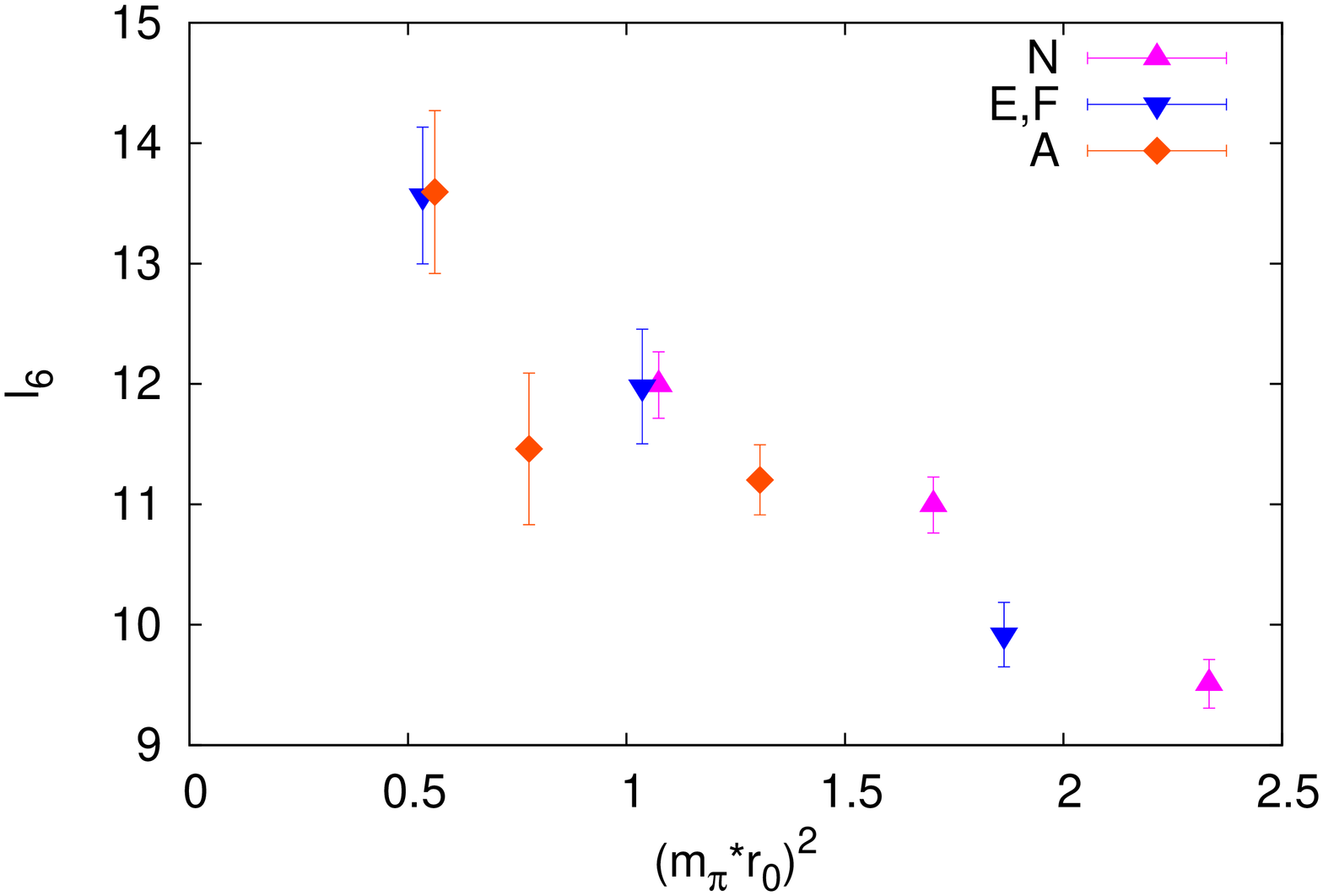}
\vspace*{-3mm}
\caption{{\bf Left:} Results for the pion charge radius extracted from linear
fits to $f_{\pi\pi}(q^2)$ in the region $(r_0\:q)^2\leq-0.15$ against
$\left(m_\pi\:r_0\right)^2$ together with results from other collaborations.
{\bf Right:} Results for $\bar{\ell}_6$ against $\left(m_\pi\:r_0\right)^2$.}
\label{fig3}
\vspace*{-3mm}
\end{figure}

The results for the pion form factor on our lightest ensemble for
each lattice spacing are shown in figure \ref{fig1}, together with the results
from \cite{Nguyen:2011ek,Frezzotti:2008dr,Boyle:2008yd} and the experimental
results from \cite{Amendolia:1986wj}. Our points reach down to
$(r_0\:q)^2\lesssim-10^{-4}$ with small statistical uncertainties. This
enables us to use a linear fit to $f_{\pi\pi}(q^2)$ in the region
$(r_0\:q)^2\leq-0.15$ to extract $\Erw{r_\pi^2}$ without any model
dependence. The results from the linear fit are shown in figure \ref{fig3}
(left) for all ensembles, together with the other results quoted above.
We see consistency with the results from other collaborations. Note
that our results might show some residual cut-off dependence, investigated in
our forthcoming publication.

Since our quark masses are bigger than the physical ones, we have to perform a
chiral extrapolation to the physical point, guided by chiral
perturbation theory ($\chi$PT). The $\chi$PT expressions for $f_{\pi\pi}$, and
thus also for $\Erw{r_\pi^2}$, have been worked out to next-to-next-to leading
order (NNLO) in \cite{Bijnens:1998fm}. As an intermediate step to compare to the
NNLO formulae we start with the comparison to NLO, which was derived in
\cite{Gasser:1983yg}. In figure \ref{fig3} (right) we show the results for the
only free parameter $\bar{\ell}_6$ from the fit with $(r_0\:q)^2\leq-0.15$
against $\left(m_\pi\:r_0\right)^2$. If $\chi$PT to NLO were a good description
for the mass range of our simulations we would expect $\bar{\ell}_6$ to be
constant, which is apparently not the case. This observation is consistent with
the findings of earlier studies as e.g. in
\cite{Nguyen:2011ek,Frezzotti:2008dr}. To make a statement on $\Erw{r_\pi^2}$ at
the physical point we thus envisage to use $\chi$PT to NNLO. In addition, finite
volume effects as well as lattice artefacts, which might still be present in our
analysis, are not taken into account so far. The discussion of the corresponding
analysis of both, $\chi$PT to NNLO and finite volume effects and lattice
artefacts, is postponed to a later publication.


{\bf Conclusions:} In this proceedings article we have given an overview on our
ongoing determination of the electromagnetic form factor of the pion in lattice
QCD. We use twisted boundary conditions to attain a high density of measurements
around $q^2=0$ which allows us to extract the charge radius without residual
model dependence. We have compared our measurements to $\chi$PT at NLO and
conclude that NLO is insufficient to describe the data in this mass range
consistently. In the final analysis we are going to compare our measurements to
$\chi$PT at NNLO and perform detailed studies on cut-off effects, finite volume
effects and contributions to the form factor from excited states.

{\it Acknowledgements:} {\small We like to thank our colleges from CLS for
sharing ensembles. The simulations were done on the dedicated QCD cluster Wilson
in Mainz. This work is funded in parts by SFB 443 of the DFG.}



%

}  



\begin{thebibliography}{99}
 
\bibitem{Durr:2008zz}
  S.~Durr {\it et al.},
  Science {\bf 322 } (2008)  1224,
  [arXiv:0906.3599 [hep-lat]].
\bibitem{Colangelo:2010et}
  G.~Colangelo {\it et al.},
  Eur.\ Phys.\ J.\  {\bf C71 } (2011)  1695,
  [arXiv:1011.4408 [hep-lat]].
\bibitem{Alexandrou:2010cm}
  C.~Alexandrou,
  PoS {\bf LATTICE2010 } (2010)  001,
  [arXiv:1011.3660 [hep-lat]].
\bibitem{Collins:2011mk}
  S.~Collins {\it et al.},
  [arXiv:1106.3580 [hep-lat]].
\bibitem{Capitani:2010sg}
  S.~Capitani, B.~Knippschild, M.~Della Morte, H.~Wittig,
  PoS {\bf LATTICE2010 } (2010)  147,
  [arXiv:1011.1358 [hep-lat]].
\bibitem{Brandt:2011sj}
  B.~B.~Brandt {\it et al.},
  arXiv:1106.1554 [hep-lat].
\bibitem{Boyle:2008rh}
  P.~A.~Boyle {\it et al.},
  JHEP {\bf 0808 } (2008)  086,
  [arXiv:0804.1501 [hep-lat]].
\bibitem{Bedaque:2004kc}
  P.~F.~Bedaque,
  Phys.\ Lett.\  {\bf B593 } (2004)  82,
  [nucl-th/0402051].
\bibitem{deDivitiis:2004kq}
  G.~M.~de Divitiis, R.~Petronzio, N.~Tantalo,
  Phys.\ Lett.\  {\bf B595 } (2004)  408,
  [hep-lat/0405002].
\bibitem{Boyle:2007wg}
  P.~A.~Boyle {\it et al.},
  JHEP {\bf 0705 } (2007)  016,
  [hep-lat/0703005 [HEP-LAT]].
\bibitem{Sommer:1993ce}
  R.~Sommer,
  Nucl.\ Phys.\  {\bf B411 } (1994)  839,
  [hep-lat/9310022].
\bibitem{Donnellan:2010mx}
  M.~Donnellan {\it et al.},
  Nucl.\ Phys.\  {\bf B849 } (2011)  45.
  [arXiv:1012.3037 [hep-lat]].
\bibitem{hip_latt2011} G. von Hippel, Talk presented at the International
Symposium on Lattice Field Theory, ``Lattice 2011''.
\bibitem{Brandt:2010ed}
  B.~B.~Brandt {\it et al.},
  PoS {\bf LATTICE2010 } (2010)  164,
  [arXiv:1010.2390 [hep-lat]].
\bibitem{Nguyen:2011ek}
  O.~H.~Nguyen, K.~-I.~Ishikawa, A.~Ukawa, N.~Ukita,
  JHEP {\bf 1104 } (2011)  122,
  [arXiv:1102.3652 [hep-lat]].
\bibitem{Frezzotti:2008dr}
  R.~Frezzotti {\it et al.} [ ETM Collaboration ],
  Phys.\ Rev.\  {\bf D79 } (2009)  074506,
  [arXiv:0812.4042 [hep-lat]].
\bibitem{Boyle:2008yd}
  P.~A.~Boyle {\it et al.},
  JHEP {\bf 0807 } (2008)  112,
  [arXiv:0804.3971 [hep-lat]].
\bibitem{Amendolia:1986wj}
  S.~R.~Amendolia {\it et al.} [ NA7 Collaboration ],
  Nucl.\ Phys.\  {\bf B277 } (1986)  168.
\bibitem{Bijnens:1998fm}
  J.~Bijnens, G.~Colangelo, P.~Talavera,
  JHEP {\bf 9805 } (1998)  014,
  [hep-ph/9805389].
\bibitem{Gasser:1983yg}
  J.~Gasser, H.~Leutwyler,
  Annals Phys.\  {\bf 158 } (1984)  142.

\end{thebibliography}
\end{document}